# Student-based Collaborative Network for Delivering Information of Natural Disasters and Climate Adaptation *

Huynh Dinh Chien, Huynh Thi Xuan Phuong, *Learning Resource Center, Hue University*

*Abstract*— The student generation nowadays is considered as the Net Generation who grow up in the age of Internet and mobile technology and prefer bringing technological application into their life. The Learning Resource Center – an academic information center - takes the initiative in developing a collaborative network with the students' participation for delivering information of natural disasters and climate adaptation to the people who live in the vulnerable areas of Thua Thien Hue province. The network is aimed at helping local people mitigating vulnerabilities and adapting to the globally climate change for a better life.

*Keywords*— collaborative network, student-based network, youthful participation, information delivery, natural disasters, climate adaptation.

## I. INTRODUCTION

Hue University locates on Thua Thien Hue province – the educational and cultural center of the Central Vietnam. Found in 1957, it nowadays becomes a high-quality center for educating and training human resource for both the province and the central region. Learning Resource Center (LRC) is an academic information center of Hue University, providing students and academic staff with information and media for learning, research and teaching method improvement. Found in 2002 under the totally financial support by The Atlantic Philanthropies, LRC is seen as a "common home" for students. LRC annually issues nearly 7,500 membership cards to clients, in which students occupy 92% of it.

LRC has built up their information resources based on user needs and wide range of disciplines. It has had more than 22,000 item titles in printed or electronic format. It has been recently developed a small-scale virtual library of medical documents such as 3D animations, video clips and images used in e-learning. Users can access nearly 5,000 virtual items through the interface of D-space open source software. LRC is a computer-based and internet-connected center for the whole Hue University's students, academic staff and researchers. Annually, LRC furthermore organizes interactive activities to connect users to collections and library.

There are approximately 13% of students who are LRC's clients living in rural areas of Thua Thien Hue province. As an effort of enhancing life quality in rural areas by the government, the youth today has had a better condition in approaching better education and new technological devices, and exploring social networking. They are called Net Generation or Millennials (Oblinger and Oblinger, 2005). They tend to be more active, confident and knowledgeable. They are likely to prefer new technologies and challenges, and taking part in widely community-affected activities. Though this description addresses to students in general, Vietnamese young students also engage in the same trend of ubiquitous Net Generation.

The youth participatory role in community-based activities is increasingly important and significant. It is the fact that some of development strategies have greater reference to young people who have stronger voice in families and communities. Vietnamese youth, for instance, have participated in Poverty Reduction Strategy by government, in which their youthful perspective can highlight the unrecognized areas leading to significant positive policy changes (DFID-CSO Youth Working Group, 2010). Young age, knowledge and responsibility are likely the motivation for them to take part in these activities.

## II. ACTUAL SITUATION

Vietnam is one of the countries in the East Asia and the Pacific heavily suffering from natural disasters and climate change. Thua Thien Hue province has a complex river and lagoon system and a dynamic coastline, where are the habitats for millions of species and mean of livings for thousands of local people. It is also one of the regions frequently attacked by hurricanes and typically faced to the high risk of sea-level rising.

One of the solutions for the above problems is to provide local people with fully information that helps them mitigating damage, building stronger resilience and adapting to the climate change. It is clearly that an informed person will be more active in crisis management. However, the way of information delivery from central to local has not been done in an effective way that local people can sufficiently receive and fully absorb (24-NQ/TW Resolution, 2013). For instance, they have still been in a passive motive in coping with an emergency problem, e.g. an epidemic disease, or they have not been aware of that their waste from increasingly aquacultural activities may break the balance of lagoon ecosystem. Delivering natural disaster early warnings to local communities has been furthermore still slow (Nguyen, 2007). The traditional ways of information transferring are mainly through official letters, mass media (TV, broadcasting, and newspapers), short message services (SMS) and other physical approaches. Technological applications have not been utilized in this area at an adequate level. Especially, an integrated disaster management data system has not been built to help local government and people for better preparedness (Nguyen, 2007). In the issue of climate adaptation, people's understanding of climate change and its long-term consequences has been still vague (24-NQ/TW Resolution, 2013). There exists a problem that a proportion of local people consider climate change as a global issue that there is no way to improve.

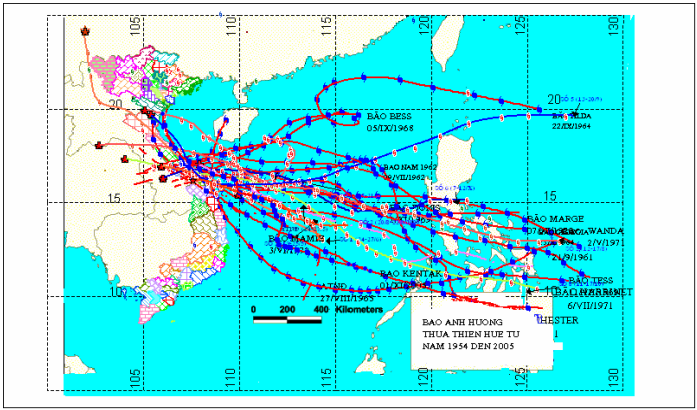

The storms attacked Thua Thien Hue province from 1954-2005 (Adapted from *Thien tai o TTH va cac bien phap phong tranh tong hop*, Nguyen, 2007)

### III. CONCEPT FORMATION

The need of community-based natural hazard management, the youthful responsibility and the Web 2.0 generation seem to be likely a meaningful connection that inspires us, especially in the situation that the existing information delivery system from central to local lacks for youth participation and technologies. In the need of bettering the traditional system, LRC proposes to build a collaborative network with the participation of students and local people who are the beneficiaries. It connects the campus to the public and takes the role as a channel of information transferring. This initiative is presented in this report as a concept paper.

### IV. NETWORK MODEL

*A. Overview*

It is intended that the network model has three levels: information center – the information provider, students – the collaborators, and local people – the information receivers or the beneficiaries. Only the communes of the coastal districts are selected because they are likely the most vulnerable and fragile regions under the attack of natural disasters and sea level rising (Tran, 2012; Nguyen, 2012).

There are 30 communes of 5 districts locating along the lagoon and the coastline to be taken into account at the first stage. Each collaborator is in charge of one commune, the number of students required is about 30. There is a priority for those who live in the chosen communes to be selected. Information receivers are selected from the households classified as "poor" or "nearly poor" in the communes who voluntarily participate. 15 households per commune is an expected number.

*B. Network Objectives*

The network is aimed at:
- Raising public awareness about natural disaster management and climate change
- Ensuring local people and the youth to be provided and updated with fully information about natural disaster management and climate change
- Assisting local people and the youth managing the risk and protecting themselves from crises, acting together against climate change.

*C. Network Significance*

- The network brings the benefit to both local people and the youth for better natural disaster preparedness and climate adaptation.
- It is a specific network as collaborators do this work for the benefit of their own family, neighbors, friends and their community, which is a good motivation for them to take part in.
- Local people would rather receive information or guidelines from their family member or the person they know than the others.
- Collaborators can keep track of attitude and awareness changing of their community after getting benefit from the network.
- It is a cost-effective and technically less-complicated approach.

*D. Network Recruitment*

Students will be recruited through CVs and interviews by LRC. They are required to have a good knowledge about environment, climate change and socio-economic issues and the communication skill. It is privileged to students who are also the local people in the sensitive communes. The winner candidates will be trained by LRC and to be certified as practitioner collaborators.

Local people who are in poor condition are selected. They are required to fill out an information sheet to provide their contact information and agree to give feedback if needed.

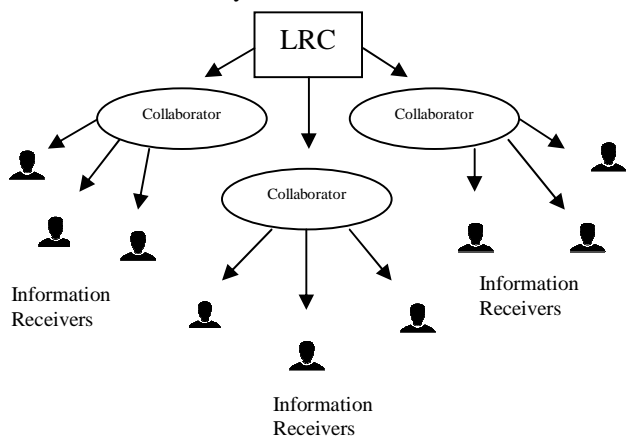

*E. Task Allocation*

Tasks are clearly allocated to three levels. As a hub of information, LRC is in charge of collecting, synthesizing information, building a data storage and access system, and prepare user interface. In responsible for information contents, it must check out its reliability, currency, copyright, relevancy before launching information to collaborators. It also takes the lead in organizing training courses for collaborators, holding three-level workshops for gathering feedbacks when needed, and supervizing the whole network.

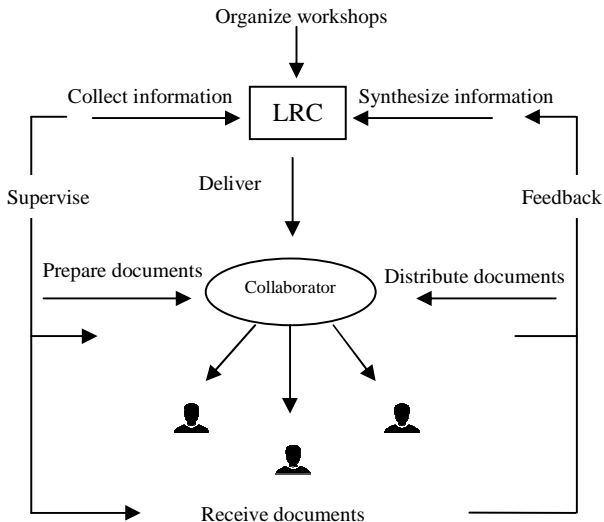

Information will be distributed to collaborators in different ways and in different formats. Collaborators will be in charge of copying documents (if in printed format), or use their devices to show (if in audio-visual format), etc. In the case of emergency, information must be delivered as soon as possible in any way (not only in paper-based format) that local people can capture easily and quickly. Collaborators are not allowed to edit or change the documents.

Local people in turn use information and give feedback. They can refuse to continue participating in the information circular. They will be asked about the usefulness of the information in the evaluation stage.

*F. Information Content and Format*

There are two kinds of information to be launched based on the using purpose:
- Regular information: raising awareness, changing behavior, guiding (manuals, guidelines, booklets, leaflets etc.)
- Emergency information: informing an emergency case, warning, alerting (massages, emails, phone calls)

Information types can be video clips, animations, guidelines, booklets, leaflets, short massages, slogans or statistics etc. Information content focuses on two issues: natural disasters management and climate change. The documents must be readily prepared for different periods:

- Before disaster (prevention, warning)
- Disaster happening (alert)
- After disaster (recovery)

Choosing a suitable format and delivery medium that local people can receive quickly and absorb effectively should be taken into account by the provider:
- LRC → collaborators: audio-visual, electronic format through emails/web upload, social networking, instant messages; voice through cell phones, home phones.
- Collaborators → local people: audio-visual, paper-based format by transferring in-person (mainly); voice through cell phones, home phones; electronic format through emails.

*G. Data Management*

Data management system is basically simple and cost-effective, as depicted in the figure below:

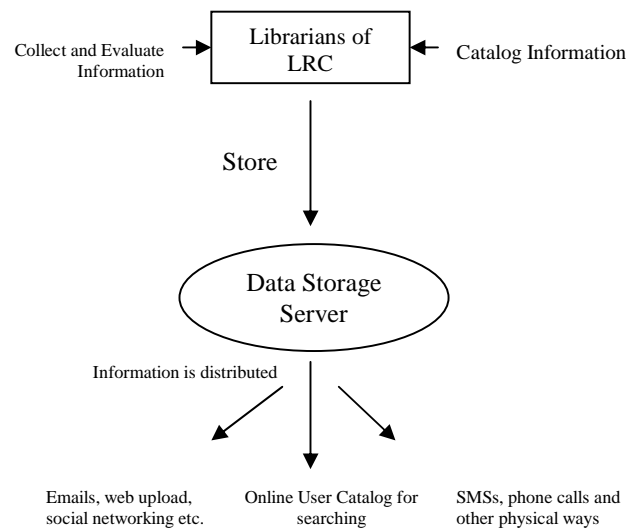

*H. Network Process*

Process of tasks and information flow is figured out in the circle below. Supervision is to examine whether collaborators fulfill their tasks. Evaluation is to checking up on the network usefulness before upgrading it to a higher level.

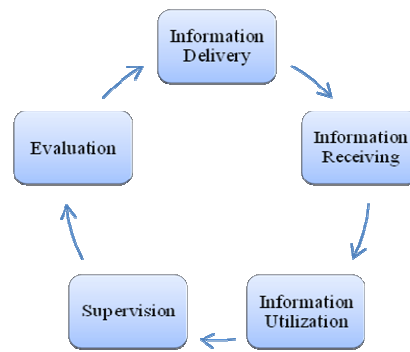

*I. Supervision*

It is needed to supervise the network operation in which collaborators will be checked out. Each collaborator will be provided with a code for management. Depending on the way of delivery, the provider will use one or more than a supervision method as follow:

- Carbon copy to LRC (emails)
- Collaborator cross-checks
- Phone calls to an information receiver at random
- Signatures of information receivers as the evidence.

*J. Network Maintenance*

One of the disadvantages of the network is students will finish their studies and graduate. Network maintenance is therefore necessarily taken into consideration. The first approach to be proposed is that one collaborator can introduce and foster one or two substitute persons who may be younger students. These younger students will be the alternative collaborators for the next generation. The second approach is to carry out follow-up activities for the purpose of adhering students to the network and community responsibility, e.g. field trips, reliefs, awards to the best collaborators.

*K. Operational Finance*

Finance is always a critical issue when this initiative is still on paper. It is desirable that the fund may come from:

- International organizations
- Government budget
- Corporate social responsibility.

## V. EVALUATION

To evaluate the network, the methods of questionnaire survey and key person interview will be used, in which both students and local people are the subjects. Usefulness of the network will be an aspect to pose questions about. Through this survey, user information needs are also measured and described in the findings. The analyzed results will be considered to decide whether the network is upgraded.

## VI. NETWORK UPGRADING

If the pilot stage is evaluated to be successful, the project will be upgraded and expanded at a higher level:

- The network will deliver training courses to local people, not only providing them with information.
- It will recruit more collaborators such as pupils and young teachers in vulnerable communes, forming a school-based network.
- LRC will build an open integrated data system on disaster management and climate change for lower levels' free access.

## VII. CONCLUSION

The student-based collaborative network is one of the channels assisting the local government to improve the existing propaganda program. It is able to figure out its outputs are (i) information is delivered more effectively from higher level to grassroots level and (ii) the network is expected to address grassroots to safer life for their own and their community. However, the network needs to be improved and collaboration of relevant agents must be strengthened. In response to the appeal to acting together because of a better disaster management strategy and against climate change, with our powerful student force and experience in information management, Hue LRC, in collaboration or cooperation with other organizations, is eager and ready to participate and may bring a good contribution into this area for safer life of local people.

REFERENCES

[1] 24-NQ/TW Resolution Chu dong ung pho voi bien doi khi hau, tang cuong quan ly tai nguyen va bao ve moi truong, *Bao Nhan dan,* Jun 2013, retrieved 5 Sept 2013 from http://www.nhandan.com.vn/phapluat/van-ban-moi/item/20496902-nghi-quyet-ve-chu-dong-ung-pho-voi-bien-doi-khi-hau-tang-cuong-quan-ly-tai-nguyen-va-bao-ve-moi-truong.html
[2] DFID-CSO Youth Working Group. "Youth Participation in Development – Aguide for Development Agencies and Policy Makers: the Summary". Restless: London, 2010, pp 17.
[3] D. H. Nguyen, "Cac nhan to anh huong den su lua cho sinh ke cua cac nong ho tai vung cat ve bien tinh Thua Thien Hue". *Tap chi Dai hoc Hue*, vol. 72B, no. 3,pp. 93-102, 2012.
[4] D. N. Oblingler, J. L. Oblinger, "Is it Age or IT: First Step Toward Understanding the Net Generation", in *Educating the Net Generation,* D. N. Oblingler, J. L. Oblinger, Washington: Educause, 2005, pp. 2.1-2.20.
[5] H. T. Tran, "Nang cao kha nang thich ung voi bien doi khi hau cho nguoi dan ven bien o tinh Thua Thien Hue", *Tap chi Dai hoc Hue*, vol. 72B, no. 3,pp. 379-386, 2012.
[6] V. Nguyen, "Thien tai o Thua Thien Hue va cac bien phap phong tranh tong hop", presented at the workshop *Xay dung nguy co song than cho cac vung bo bien Viet Nam,* Hue, 2007. Retrieved 5 Sept 2013 from http://www.imh.ac.vn/b_tintuc_sukien/bc_hoinghi_hoithao/L500-thumuccuoi/.